\documentclass[epjCONF]{svjour}
\usepackage{graphics}
\usepackage[varg]{txfonts} 
\usepackage[latin1]{inputenc}
\session-title{MESON2012 - 12th International Workshop on Meson
Production, Properties and Interaction}
\begin{document}
\title{Chiral dynamics with vector fields: an application to $\pi\pi$ and $\pi K$ scattering}
\author{I.V. Danilkin\inst{1}\inst{2}\fnmsep\thanks{\email{i.danilkin@gsi.de}} \and M.F.M. Lutz\inst{1}}
\institute{GSI Helmholtzzentrum f\"ur Schwerionenforschung GmbH,\\
Planck Str. 1, 64291 Darmstadt, Germany \and SSC RF ITEP, Bolshaya
Cheremushkinskaya 25, 117218 Moscow, Russia}
\abstract{A theoretical study of Goldstone boson scattering based
on the chiral Lagrangian with vector meson fields is presented. In application of
a recently developed novel approach we extrapolate subthreshold partial-wave amplitudes
into the physical region. The constraints set by micro-causality and coupled-channel unitarity
are kept rigourously. It is shown that already the leading order subthreshold amplitudes
lead to s- and p-wave $\pi\pi$ and $\pi K$ phase
shifts are in agreement with the experimental data up to about 1.2 GeV.}
%
\maketitle
\section{Introduction}
\label{sec:Introduction}

Chiral perturbation theory ($\chi PT$) is a powerful tool for a
systematic study of the interaction of Goldstone bosons
\cite{Weinberg:1967tq,Gasser:1983yg}. However the convergence of
the chiral expansion is limited to the threshold region
and a generalization to higher energy is desirable. It was shown that
coupled-channel unitarity techniques may lead to a systematic description of
scattering phases in the resonance region \cite{GomezNicola:2001as}.

The purpose of this contribution is to apply a recently proposed
unitarization scheme \cite{Gasparyan:2010xz} to Goldstone boson
scattering based on the chiral Lagrangian supplemented with light
vector mesons. There are several reasons for considering the light
vector mesons as explicit degrees of freedom. First of all we
recall a resonance saturation mechanism \cite{Ecker:1988te}. It
was shown that the size of the low energy constants (LECs) of
$Q^4$ counter terms are basically saturated by the light vector
mesons. Second, the light vector mesons play a particular role in
the hadrogenesis conjecture
\cite{Lutz:2008km,Lutz:2003fm,Terschlusen:2012xw}. Together with
the Goldstone bosons, they are identified to be the relevant
degrees of freedom that are expected to generate the meson
spectrum. For instance it was shown that the leading chiral
interaction of Goldstone bosons with the light vector mesons
generates an axial-vector meson spectrum that is quite close to
the empirical one \cite{Lutz:2003fm}.

The Chiral Lagrangian at order $Q^2$ includes two relevant and known
parameters only, the chiral limit value of the pion decay constant
and a coupling constant that characterize the decay of the rho
meson into a pair of pions. As a result we recover the empirical
pion-pion and pion-kaon scattering up to about
$\sqrt{s}\simeq1.2$ GeV \cite{Danilkin:2011fz}.

\section{Description of the method}
\label{sec:Description of the method}

\begin{figure}[t]
\resizebox{0.75\columnwidth}{!}{\includegraphics{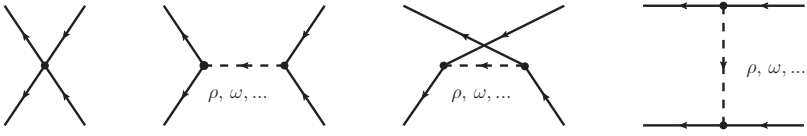}}
\caption{\label{fig.Diagrams}Tree level diagrams for Goldstone
boson scattering with the exchange of light vector mesons (dashed
line) in the s-,t- and u-channels.}
\end{figure}

The relevant leading-order Lagrangian
\cite{Lutz:2008km,Terschlusen:2012xw} is given by
\begin{eqnarray}\label{Lagrangian}
{\cal L}&=&\frac{1}{48f^2}\,{\rm tr}
\,\Big\{[\Phi,\,\partial^{\mu}\Phi]_-\,[\Phi,\,\partial_{\mu}\Phi]_-+\Phi
^4\,\chi_{0}\Big\} -i\,\frac{f_{V}\,h_{P}}{8f^2}\,{\rm tr}
\,\Big\{\partial_{\mu}\Phi\,
\Phi^{\mu\nu}\,\partial_{\nu}\Phi\Big\}\,,\nonumber
\end{eqnarray}
where the pseudo-scalar and vector mesons are collected in
\begin{eqnarray}
\Phi&=&\left(\begin{array}{ccc}
\pi^0+\frac{1}{\sqrt{3}}\,\eta &\sqrt{2}\,\pi^+&\sqrt{2}\,K^+\\
\sqrt{2}\,\pi^-&-\pi^0+\frac{1}{\sqrt{3}}\,\eta&\sqrt{2}\,K^0\\
\sqrt{2}\,K^- &\sqrt{2}\,\bar{K}^0&-\frac{2}{\sqrt{3}}\,\eta
\end{array}\right)\quad\textrm{and}\quad
\Phi_{\mu \nu} = \left(\begin{array}{ccc}
\rho^0_{\mu \nu}+\omega_{\mu \nu} &\sqrt{2}\,\rho_{\mu \nu}^+&\sqrt{2}\,K_{\mu \nu}^+\\
\sqrt{2}\,\rho_{\mu \nu}^-&-\rho_{\mu \nu}^0+\omega_{\mu \nu}&\sqrt{2}\,K_{\mu \nu}^0\\
\sqrt{2}\,K_{\mu \nu}^- &\sqrt{2}\,\bar{K}_{\mu
\nu}^0&\sqrt{2}\,\phi_{\mu \nu}
\end{array}\right)\,,
\end{eqnarray}
respectively. In order to generate a faithful resonance saturation
mechanism, the vector mesons are represented in terms of
anti-symmetric fields $\Phi_{\mu\nu}=-\Phi_{\nu\mu}$. In the
Lagrangian (\ref{Lagrangian}), $f=90$ MeV may be identified with
the pion-decay constant at leading order. The value of the
parameter $f_{V}\,h_{P}$ was determined in \cite{Lutz:2008km},
\begin{equation}\label{}
f_{V}\,h_{P}\approx 0.23~~\textrm{GeV}\,.
\end{equation}
The tree level p.w. scattering amplitudes are given by the
Weinberg-Tomozawa and the $s$-, $t$- and $u$-channel vector meson
exchange terms. In Fig.\ref{fig.Diagrams} the set of diagrams that
we take into account is depicted.

\begin{figure}[b]
\resizebox{0.90\columnwidth}{!}{\includegraphics{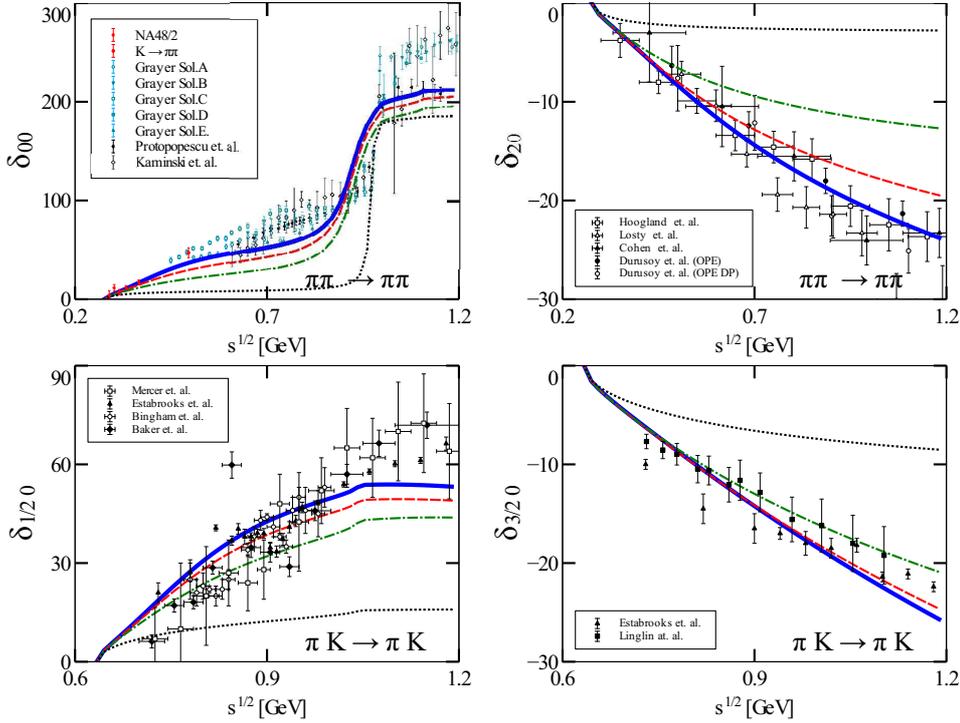}}
\caption{\label{fig.truncation order}Results for the s-wave
$\pi\pi$ and $\pi K$ phase shifts $\delta_{IJ}$. The dotted,
dash-dotted, dashed and solid lines correspond to a truncation in
the expansion (\ref{U expansion}) at order $0,\ 1,\ 2,\ 3$
respectively. For data references see \cite{Danilkin:2011fz}.}
\end{figure}

Our approach is based on partial-wave dispersion relations.
Following \cite{Gasparyan:2010xz} a generalized potential
$U^J_{ab}(s)$ is constructed from the chiral Lagrangian in the
subthreshold region and analytically extrapolated to higher
energies. The partial-wave scattering amplitudes  $T^J_{ab}(s)$
are obtained as solutions of the non-linear integral equation
\begin{eqnarray}\label{def-non-linear}
T^J_{ab}(s)&=&U^J_{ab}(s)+\sum_{c,d}\int^{\infty}_{\mu_{thr}^2}\frac{d\bar{s}}{\pi}\frac{s-\mu_M^2}{\bar{s}-\mu_M^2}\,
\frac{T^J_{ac}(\bar s)\,\rho^J_{cd}(\bar s)\,T^{J*}_{db}(\bar
s)}{\bar{s}-s-i\epsilon}\,,
\end{eqnarray}
which leads to a controlled realization of the causality and
coupled-channel unitarity condition. Here $\rho^J_{cd}(s)$ is the
phase-space matrix and $\mu_M$ is a matching scale which is
identified with the smallest two-body threshold accessible in a
sector with isospin and strangeness $(I,S)$.

\begin{figure}[t]
\resizebox{0.90\columnwidth}{!}{\includegraphics{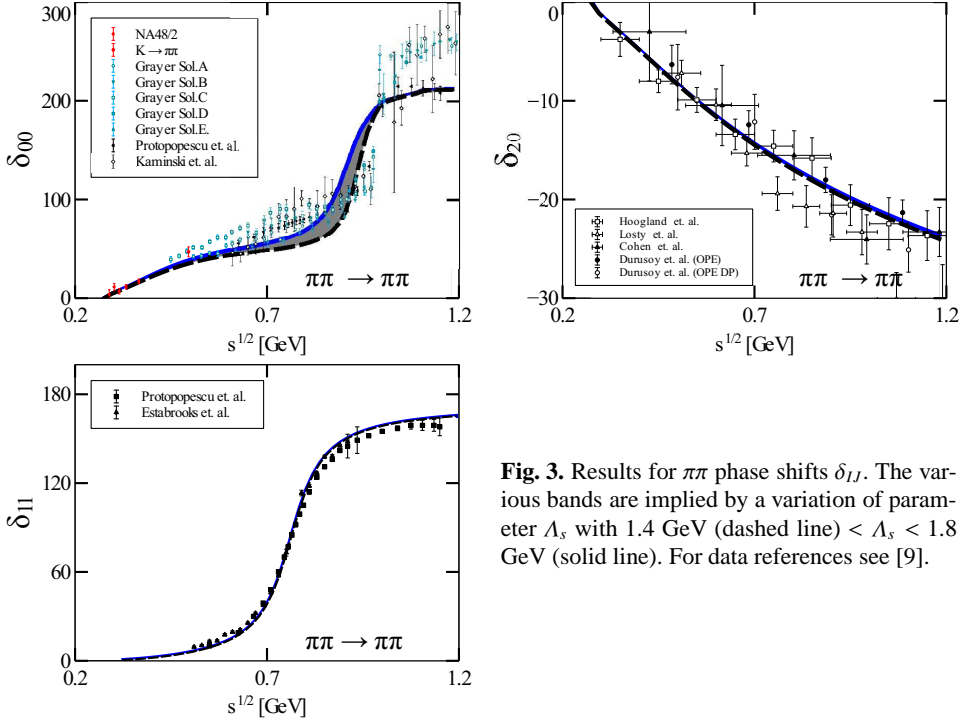}}
\parbox{11.0cm}{\vspace{-6.5cm}
\hspace*{6.5cm} \parbox{6cm}{\caption{\label{fig:pipi}Results for
$\pi\pi$ phase shifts $\delta_{IJ}$. The various bands are implied
by a variation of parameter $\Lambda_s$ with 1.4 GeV (dashed line)
$< \Lambda_s <$ 1.8 GeV (solid line). For data references see
\cite{Danilkin:2011fz}.}}}
\end{figure}

As was pointed out in \cite{Gasparyan:2010xz}, the generalized
potential $U_{ab}^J(s)$ can be reconstructed unambiguously in
terms of its derivatives at an expansion point that lies within
its analyticity domain and where the results of $\chi$PT are
reliable. It holds
\begin{eqnarray}\label{U expansion}
U(s)=\sum_{k=0}^{n} \,c_k \,\xi^k(s)\quad \textrm{for}\quad s <
\Lambda_s^2 \,,
\end{eqnarray}
where the coefficients $c_k$ are determined by the first $k$
derivatives of $U(s)$ at the expansion point. The function
$\xi(s)$ is a suitable conformal map constructed as to
analytically continue the potential to larger energies.
Explicitly, it is given by
\begin{eqnarray} \label{def-conformal}
\xi(s)=\frac{a\,(\Lambda^2_s-s)^2-1}{(a-2\,b)(\Lambda^2_s-s)^2+1}\,,\quad
a = \frac{1}{(\Lambda^2_s-\mu_E^2)^2},
 \quad b = \frac{1}{(\Lambda^2_s-\Lambda^2_0)^2}\,,
\end{eqnarray}
where the parameter $\Lambda_0$ is identified unambiguously such
that the mapping domain of the conformal map touches the closest
left-hand branch point. The value of $\Lambda_s$ sets the scale
from where on s-channel physics is integrated out.  For
$s>\Lambda_s^2$ the generalized potential is set to a constant.

In the p-wave amplitude, vector mesons show up as poles above
threshold. In that case CDD poles have to be included explicitly
in order to solve Eq.(\ref{def-non-linear}).

\section{Results}

\label{sec:Results}

We first study the dependence of our results  on the truncation index $n$ in (\ref{U expansion}) at
given $\Lambda_s=1.6$ GeV. In Fig.\ref{fig.truncation order} the s-wave phases shifts for $\pi \pi$ and $\pi \,K$
are shown for the cases $n=0,1,2,3$. The strong dependence on the truncation index we interpret as a clear signal of
vector meson dynamics in the s-wave phase shifts. The convergence pattern towards the empirical phase shifts is quite reassuring pointing at the reliability of our approach. We note that in the p-wave scattering phases there is a very minor dependence on the
truncation index $n$ only.

\begin{figure}[t]
\resizebox{0.90\columnwidth}{!}{\includegraphics{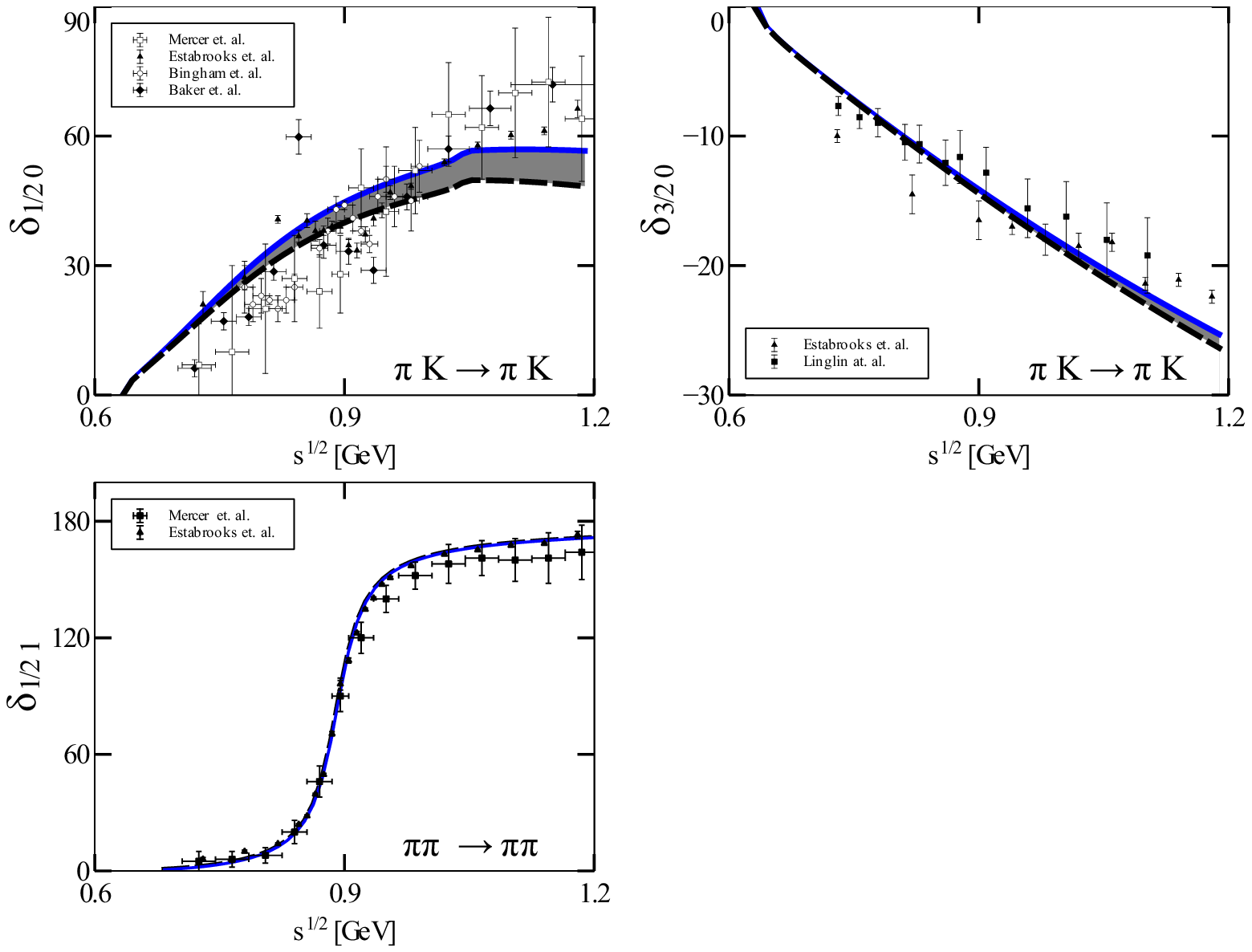}}
\parbox{11.0cm}{\vspace{-6.5cm}
\hspace*{6.5cm} \parbox{6cm}{\caption{\label{fig:piK}Results for
$\pi K$ phase shifts $\delta_{IJ}$. The various bands are implied
by a variation of parameter $\Lambda_s$ with 1.4 GeV (dashed line)
$< \Lambda_s <$ 1.8 GeV (solid line). For data references see
\cite{Danilkin:2011fz}.}}}
\end{figure}

In Fig. \ref{fig:pipi} and \ref{fig:piK}  we show the dependence of the scattering phases
for fixed $n=3$ but a  variation of $\Lambda_s$ from 1.4 GeV to 1.8 GeV. In all cases this causes
a rather small error band only, where for $n<3$ the error bands get even smaller.
We conclude that within our scheme the s- and p-wave $\pi\pi$ and $\pi K$ phase shifts are in a good
agreement with the data set up to 1.2 GeV.

\end{document}